\documentclass[pre,showpacs,floats,twocolumn,floatfix,superscriptaddress]{revtex4-1}

\usepackage{graphics,graphicx,dcolumn,bm,fleqn,epic,eepic,float}
\usepackage{amssymb,amsmath,multirow,rotate,rotating,color}
\usepackage[utf8]{inputenc}

\definecolor{tuered}{RGB}{214,0,74}
\definecolor{tueblue}{RGB}{0,102,204}
\definecolor{grey}{RGB}{128,128,128}

\renewcommand{\vec}[1]{\mathbf{#1}}

\usepackage[]{cleveref}

\newcommand{\cfig}[1]{\centering\includegraphics[width=0.9\linewidth]{#1}} 
\newcommand{\etal}{\emph{et al}} 

\newcommand{\den}{\rho}

\newcommand{\Pos}{\vec{x}}

\renewcommand{\time}{t}

\newcommand{\svol}{V_{\mathrm{sim}}}

\newcommand{\ffr}{\chi}

\newcommand{\op}{\varphi}

\newcommand{\st}{\sigma}

\newcommand{\pc}{\Delta\rho}

\newcommand{\pca}{\theta_p}

\newcommand{\pr}{r_p}

\newcommand{\dr}{R_d}

\newcommand{\alds}{\Lambda}

\newcommand{\pvf}{\Xi}

\begin{document}

\title{Domain and droplet sizes in emulsions stabilized by colloidal particles}

\author{Stefan Frijters}
\email{s.c.j.frijters@tue.nl}
\affiliation{Department of Applied Physics, Eindhoven University of Technology, Den Dolech 2, NL-5600MB Eindhoven, The Netherlands}

\author{Florian G\"unther}
\email{f.s.guenther@tue.nl}
\affiliation{Department of Applied Physics, Eindhoven University of Technology, Den Dolech 2, NL-5600MB Eindhoven, The Netherlands}

\author{Jens Harting} 
\email{j.harting@tue.nl}
\affiliation{Department of Applied Physics, Eindhoven University of Technology, Den Dolech 2, NL-5600MB Eindhoven, The Netherlands}
\affiliation{Faculty of Science and Technology, Mesa+ Institute, University of Twente, 7500 AE Enschede, The Netherlands}

\date{\today}

\begin{abstract}
Particle-stabilized emulsions are commonly used in various industrial
applications. These emulsions can present in different forms, such as Pickering
emulsions or bijels, which can be distinguished by their different topologies
and rheology. We numerically investigate the effect of the volume fraction and
the uniform wettability of the stabilizing spherical particles in mixtures of two fluids. For
this, we use the well-established three-dimensional lattice Boltzmann method,
extended to allow for the added colloidal particles with non-neutral wetting
properties. We obtain data on the domain sizes in the emulsions by using both
structure functions and the Hoshen-Kopelman (HK) algorithm, and demonstrate
that both methods have their own (dis-)advantages. We confirm an inverse
dependence between the concentration of particles and the average radius of the
stabilized droplets. Furthermore, we demonstrate the effect of particles
detaching from interfaces on the emulsion properties and domain size
measurements.
\end{abstract}


\maketitle

\section{Introduction}
\label{sec:emulsions-intro}

In recent decades, many branches of industry have started using colloidal particles
to stabilize emulsions for various purposes, such as cosmetics, improved
low-fat food products, and drug delivery~\cite{ bib:dickinson:2010,
bib:pink-shajahan-razul:2014}. Emulsions are also extremely relevant to the
petroleum industry: they are commonly found during the production of crude oil,
the extraction of bitumen from tar sands, and in many other related
processes~\cite{ bib:tambe-sharma:1994}. 

Traditionally, amphiphilic surfactant molecules are often employed as emulsification agents, but their effects can be mimicked or supplemented by the use of particles. These may fill the role of a cheaper or less toxic alternative to surfactants, or they may be customized to include additional desirable properties. Examples include ferromagnetic particles~\cite{ bib:melle-lask-fuller:2005, bib:kim-stratford-cates:2010} (which can be detected remotely), Janus particles~\cite{ bib:binks-fletcher:2001, bib:glaser-adams-boeker-krausch:2006, bib:aveyard:2012, bib:kumar-bumjun-tu-lee:2013} (having different interfacial properties on different parts of their surface, improving their stabilization properties), or particles with nonspherical geometries~\cite{ bib:hore-laradji:2008, bib:zhou-cao-lui-stoyanov:2009, bib:grzelczak-vermant-furst-lizmarzan:2010, bib:botto-lewandowski-cavallaro-stebe:2012, bib:guenther-frijters-harting:2014} (which introduce additional geometrical degrees of freedom and also exhibit improved stabilization properties).

Due their amphiphilic ineractions with the involved fluids, surfactants reduce the interfacial tension, which reduces the total interfacial free energy. By contrast, colloidal particles stabilize emulsions kinetically because they can reduce interfacial free energy by replacing energetically expensive fluid-fluid interfacial area by cheaper particle-fluid interfacial area when a particle adsorps to such a fluid-fluid interface. These effects are described in some detail in~\cite{ bib:frijters-guenther-harting:2012}. 
The energy differences involved in the case of colloidal particles are generally orders of magnitude larger than thermal fluctuations, which makes this adsorption process practically irreversible~\cite{ bib:binks:2002, bib:davies-krueger-coveney-harting:2014}. Once adsorped, the particles block Ostwald ripening~\cite{ bib:ostwald:1901, bib:vengrenovitch:1982, bib:enomoto-kawasaki-tokuyama:1987, bib:voorhees:1992, bib:taylor:1998}, which is one of the main processes leading to drop coarsening and the eventual demixing of emulsions. Thus, by blocking this effect, particles allow for long-term stabilization of an emulsion.

Particle-stabilized emulsions can present in various forms, classified by the shapes and sizes of their fluid domains. The form that most resembles a traditional, surfactant-stabilized emulsion is the ``Pickering emulsion''~\cite{ bib:ramsden:1903, bib:pickering:1907}, which consists of particle-covered droplets of one fluid suspended in another fluid. A more recent discovery is the bicontinuous interfacially jammed emulsion gel, or ``bijel'', which is characterized by two large continuous fluid domains that are intertwined and are only stable because of the particles present at their interfaces. This form was first predicted by numerical simulations~\cite{ bib:stratford-adhikari-pagonabarraga-desplat-cates:2005}, and shortly thereafter confirmed experimentally~\cite{ bib:herzig-white-schofield-poon-clegg:2007, bib:clegg-herzig-schofield-egelhaalf-horozov-binks-cates-poon:2007}. Parameters that affect the final state of an emulsion include the ratio between the two fluid components, the volume fraction of the particles, and their wettability. These parameters have been studied numerically by various authors~\cite{ bib:kim-stratford-adhikari-cates:2008, bib:jansen-harting:2011, bib:aland-lowengrub-voigt:2011, bib:guenther-janoschek-frijters-harting:2013, bib:guenther-frijters-harting:2014}. In this publication, we expand on the work of Jansen and Harting~\cite{ bib:jansen-harting:2011} in particular and create a link with the experimental and theoretical work by Arditty \etal.~\cite{ bib:arditty-whitby-binks-schmitt-lealcalderon:2003}. 

Following Jansen and Harting~\cite{ bib:jansen-harting:2011} we use the lattice Boltzmann (LB) method, coupled to solid particles whose inter-particle interactions are simulated by molecular dynamics, to study particle-stabilized emulsions. The method is briefly explained in Sec.~\ref{sec:sim-method}. In Sec.~\ref{sec:emulsions-states} we explain in detail how to characterize the different types of particle-stabilized emulsions, introducing two ways to quantify domain sizes.
Sec.~\ref{sec:emulsions-initial} aims to give an overview of the parameter space that determines how the emulsions present, based on the volume fraction occupied by the particles, their contact angle, and the ratio between the two fluids. After making the distinction between bijels and Pickering emulsions, the (individual) domain sizes are studied in some detail. Finally, in Sec.~\ref{sec:emulsions-conclusions} we draw our conclusions and provide an outlook on how to further improve the understanding of these complex systems.

\section{Simulation method}
\label{sec:sim-method}

To model the systems in this work we primarily use the lattice Boltzmann (LB) method. This alternative to traditional Navier-Stokes solvers has proven itself to be a very successful tool for modeling fluids~\cite{bib:benzi-succi-vergassola:1992, bib:succi:2001}. The LB method allows for easy implementation of complex boundary conditions and is well-suited for implementation on parallel supercomputers, because of the high degree of locality of the algorithm~\cite{ bib:harting-harvey-chin-venturoli-coveney:2005, bib:guenther-janoschek-frijters-harting:2013}. This allows for detailed simulations of systems on the mesoscale. The method is based on the Boltzmann equation, with its positions $\vec{x}$ discretized in space on a lattice with lattice constant $\Delta x$ and with its time $t$ discretized with a timestep $\Delta t$, which we define to be $\Delta x = \Delta t \equiv 1$ for clarity. We use the Bhatnagar-Gross-Krook (BGK) collision operator~\cite{ bib:bhatnagar-gross-krook:1954} and a D3Q19 lattice, which is to say the lattice is three-dimensional and a set of velocities $\vec{c}_i$ with $i=1,\ldots,19$ connect a lattice site with itself, its nearest neighbours and next-nearest neighbours on the lattice.

The fluid-fluid interactions are described by an interaction force, calculated locally according to the approach of Shan and Chen, which allows us to model a surface tension $\st$~\cite{ bib:shan-chen:1993}. We refer to the two fluid species as ``red'' fluid ($r$) and ``blue'' fluid ($b$), respectively. The ratio of red and blue fluid present in the system is denoted $\ffr = \sum_{\vec{x}} \den^r(\vec{x}) / \sum_{\vec{x}} \den^b(\vec{x})$. To simplify statements about the local fluid composition on lattice sites, we also introduce an order parameter $\op(\vec{x},t) = \den^r(\vec{x},t) - \den^b(\vec{x},t)$, referred to as ``colour''.

We model the movement of the colloidal particles using molecular dynamics techniques and treat their coupling to the fluids by means of a modified bounce-back boundary condition as pioneered by Ladd and Aidun~\cite{ bib:aidun-lu-ding:1998,bib:ladd-verberg:2001}. Furthermore, we define a parameter $\pc$, called the ``particle colour'', which allows to control the contact angle $\pca$ of the particle by introducing a local modification to the Shan-Chen force. In particular, it has been found that the dependence of the contact angle on the particle colour can be fitted by a positive linear relation, where the value of the slope depends on the actual simulation parameters~\cite{ bib:jansen-harting:2011}. A particle colour $\pc = 0$ corresponds to a contact angle of $\pca = 90^{\circ}$, i.e. a neutrally wetting particle. A positive particle colour $\pc > 0$ leads to a contact angle $\pca > 90^{\circ}$ and the particle will be called hydrophobic (therefore equating the blue fluid to a water-like fluid and the red fluid to an oil-like fluid). Conversely, particles with $\pc < 0$ are hydrophilic and have a contact angle $\pca < 90^{\circ}$. For more details on the techniques used in the simulations the reader is referred to~\cite{ bib:frijters-guenther-harting:2012}.

\section{Characterization of emulsion states}
\label{sec:emulsions-states}

\begin{figure*}
  \centering

  \begin{tabular}{l l l }
  \includegraphics[width=0.3\linewidth]{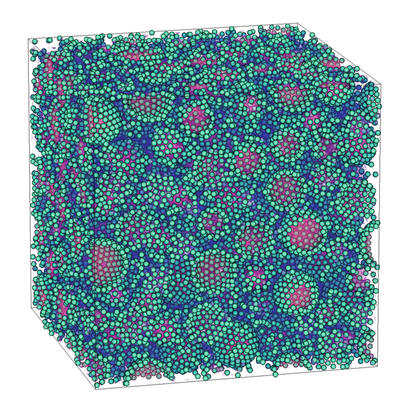} &
  \includegraphics[width=0.3\linewidth]{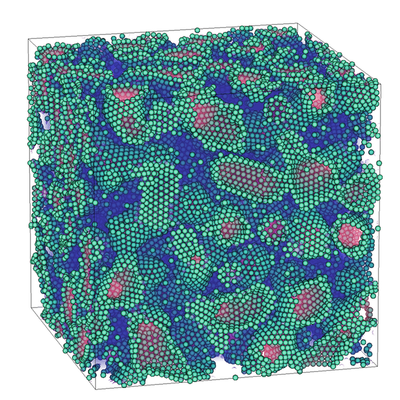} &
  \includegraphics[width=0.3\linewidth]{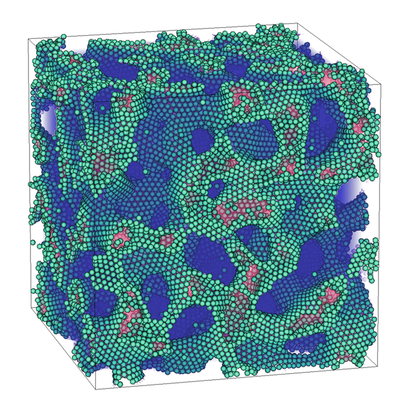} \\
  \end{tabular}
  \caption{(Colour online) Examples of emulsion states stabilized by spherical particles: The left picture showcases a Pickering emulsion (hydrophilic particles; $\pca = 77^\circ$): discrete droplets (red) suspended in the medium (blue) and stabilized by particles (green). The centre picture shows an intermediate state between a Pickering emulsion and a bijel (neutrally wetting particles $\pca = 90^\circ$): a large continuous red domain has formed, but some discrete droplets remain. Finally, the right picture depicts a bijel (hydrophobic particles; $\pca = 103^\circ$): two intertwined continuous domains.}
  \label{fig:emulsion-states}
\end{figure*}

The stability of particle-stabilized emulsions is influenced by many factors, such as the volume fraction of the particles, their size, wettability and interactions, and environmental factors, such as the ambient pH values and presence of ions~\cite{ bib:tambe-sharma:1994}. A minimum surface coverage is necessary to form a stable emulsion; this puts a lower bound on the volume fraction of the particles necessary to stabilize an emulsion. This effect is not linear, and an upper bound also exists, after which the stability of the emulsion no longer increases. The fluid-fluid ratio plays a dominant role in determining whether an oil-in-water or water-in-oil emulsion will be formed. At a fluid-fluid ratio $\ffr = 1$, the wettability of the particles is extremely important, as the curvature induced by these particles will be the deciding factor. Changing the wettability in such systems can lead to a catastrophic phase inversion~\cite{ bib:kralchevsky-ivanov-ananthapadmanabhan-lips:2005}. The size of the individual particles is also one of the most important factors when stabilizing an emulsion, because it controls the ability of a particle to reside at an interface, with smaller particles finding it easier to arrange themselves. However, as particles become smaller, the energy barrier binding them to the interface becomes small when compared to the energy of Brownian motion, such that very small particles do a poor job of stabilizing emulsions~\cite{ bib:bechhold-dede-reiner:1921}. Particle-particle interactions and electrostatic interactions can affect the denseness of the particle layers; if these cannot be sufficiently closely packed, the emulsion will not be stabilized. The presence of ions in the fluids can affect the screening length of the electrostatic interactions, and as such will dampen droplet-droplet repulsion, allowing for easier coalescence and destabilization.

Particle-stabilized emulsions can take various forms that can be characterized by the size and shape of the fluid domains. Qualitatively, we can make the distinction between Pickering emulsions, which consist of discrete droplets of one fluid inside a medium fluid, and bijels, in which each fluid forms a single continuous domain that spans the system. The transition between these phases is not sharp: intermediate states exist, in which large domains span the system, but some single droplets remain. Examples of these three states are shown in Fig.~\ref{fig:emulsion-states}. 

Jansen and Harting have previously presented phase diagrams that show the effect of the particle volume fraction and wettability as well as the effect of the fluid-fluid ratio on the emulsion state that is finally reached~\cite{ bib:jansen-harting:2011}; in this work we focus on studying different ways to characterize emulsions and restrict ourselves to modifying the volume fraction and wettability of the particles but keep the fluid-fluid ratio, particle sizes, and surface tension fixed. In particular, we use an oil-water ratio of $\ffr = 0.56$ (far away from the possibility of inverting a Pickering emulsion, but allowing bijel configurations), hard spherical particles with a radius of $\pr = 5.0$ (which balances the size of the particles in terms to the interfacial thickness of the Shan-Chen method with computational efficiency~\cite{ bib:jansen-harting:2011, bib:frijters-guenther-harting:2012}), and a surface tension of $\st = 0.014$ (which is large enough to let the fluids demix and form an emulsion).  Electrostatic effects are also not treated in this work. These choices leave us with a sufficiently large, but still manageable, parameter space in which to search for the differences between bijels and Pickering emulsions, and transitions from one to the other. To characterize the emulsions, we focus on the average sizes of the structures in the emulsions, and the individual sizes of Pickering droplets. 


In previous work, domain sizes have been calculated using the structure function of the order parameter $\op$, which, when suitably averaged, can be reduced to a time-dependent measure of the global lateral domain size $\alds$~\cite{ bib:kim-stratford-adhikari-cates:2008, bib:jansen-harting:2011, bib:guenther-janoschek-frijters-harting:2013, bib:guenther-frijters-harting:2014}. This method matches the commonly used experimental way of treating measurements of scattering data; however,
it is not without its flaws in the context of our simulations. In particular, as the colour of the particles is linked to their wettability, they will generally not blend into their environment from the point of view of this method. As such, they introduce additional length scales of the particle radius and inter-particle distance, which will depress the average value in a non-physical way (as the physical length scales of the order of the average droplet size are much larger than the radius of the particles). For details the reader is referred to~\cite{ bib:jansen-harting:2011}.


As an alternative way to measure domain sizes we use a three-dimensional variation of the Hoshen-Kopelman algorithm to detect clusters and to measure the sizes of individual fluid domains~\cite{ bib:hoshen-kopelman:1976, bib:frijters-krueger-harting:2014}. The algorithm was developed to be applied to quantities on two- or three-dimensional latticesi. It is based on detecting connected clusters on a lattice and labelling the involved lattice sites such that all sites that are connected share the same label. The assumption of the presence of a lattice makes it easy to adapt the algorithm for use in combination with LB-based simulations. Without loss of generality we assume $\ffr \le 1$, which then is a red-in-blue emulsion (the details of how we initialize the fluid mixtures will be explained in Fig.~\ref{sec:emulsions-initial}). We choose $\op(\Pos) > 0$ as the condition needed to decide whether a lattice site is part of a cluster, or part of the medium. This corresponds to the sizes of the red domains being of interest and the blue domain forming the medium. When all clusters have been identified, it is trivial to get information on domain sizes by simply counting the number of sites belonging to a particular cluster for all clusters. In addition, this method provides information on the distribution of the sizes of individual clusters (which the structure function method does not supply). A Pickering-like emulsion is characterized by a large number of small clusters (discrete droplets), while a bijel contains only a single connected cluster, which spans the system. Following Hoshen and Kopelman, we introduce the ``reduced averaged cluster size''~\cite{ bib:hoshen-kopelman:1976}
\begin{equation}
  \label{eq:racs}
  I_{\mathrm{av}} = \left( \sum_{n = 1}^{n_{\mathrm{max}}} i_n n^2 \right) / N_c - n_{\mathrm{max}}^2 / N_c
  \hbox{,}
\end{equation}
where $N_c$ denotes the total number of sites with $\op(\Pos) > 0$, $i_n$ is the number of clusters of size $n$ cubed lattice units, and $n_{\mathrm{max}}$ is the size of the largest such cluster. By thus subtracting the contribution of the single largest cluster, we observe a very sharp decrease of this quantity when transitioning from the Pickering to bijel regime. In the case of a pure bijel $I_{\mathrm{av}} = 0$, while for a Pickering emulsion $I_{\mathrm{av}} \gg 0$ contains information about the sizes of the droplets. This property makes it easy to detect a transition, but also removes all useful information about the structure of a bijel state. Even if the effect of the largest (and only) cluster would not be dropped, we only gain information that is a direct consequence of the fluid-fluid ratio and surface tension, i.e. the total volume of the red fluid. Thus, this method is best used for Pickering emulsions. Finally, in the context of Pickering emulsions we introduce a cutoff to discount unphysically small domains: due to the diffuse interface and small fluctuations, domains consisting of only a few lattice sites cannot be properly called droplets. In this work the cutoff is chosen to be 50 lattice sites, which we have found to discard most errors while retaining as much valid data as possible.

\section{Results}
\label{sec:emulsions-initial}

In this section, we discuss how changing the particle volume concentration and particle contact angle can affect the final state of a mixture of immiscible fluids and particles. Systems of various sizes are used, in order to find a system size that balances computational effort with statistical accuracy. Lower limits on system size exist, as particles have to be larger than the interfacial thickness, and multiple particle-covered droplets have to fit in the system to simulate an emulsion. As such, the simulation volumes we use consist of $\svol = 512^3$ or $\svol = 256^3$ lattice sites, with periodic boundary conditions applied on all sides. The system is initialized with red and blue fluid densities at any site chosen uniformly at random from the interval $[ 0 , \den^r_{\mathrm{init}} ]$ and $[ 0 , \den^b_{\mathrm{init}} ]$, respectively. The particles are randomly placed in the system, taking care not to allow any overlapping particles. When evolved in time, the fluids spinodally demix, and particles are swept to the newly-formed interfaces and stabilize them. This causes a rapid change of the system at early times, while at later times the system will still develop, albeit more slowly~\cite{bib:jansen-harting:2011,bib:guenther-frijters-harting:2014}. As mentioned in the previous section, we restrict ourselves to a single value of the fluid-fluid ratio $\ffr = 0.56$ ($\den^r_{\mathrm{init}} = 0.5$ and $\den^b_{\mathrm{init}} = 0.9$). Furthermore, we focus on particle volume fractions $\pvf = 0.15, 0.20, 0.25$. In this regime, the wettability of the particles has a strong effect on the final state of the emulsion. For hydrophilic particles ($\pca < 90^\circ$), the particles induce curvature favourable to a red-in-blue emulsion and droplets are formed. Therefore, a Pickering state is observed ($I_{\mathrm{av}} \gg 0$). Conversely, for hydrophobic particles ($\pca > 90^\circ$), the favoured curvature is inverted. However, there is not enough red fluid to be able to suspend blue droplets in it. Therefore, the system instead forms a bijel ($I_{\mathrm{av}} = 0$). Increasing the volume fraction $\pvf$ of the particles increases the interfacial area that they stabilize, which allows for smaller droplets in the Pickering regime.

\subsection{Domain sizes}
\label{ssec:emulsion-initial-domain-sizes}

\begin{figure}
  \cfig{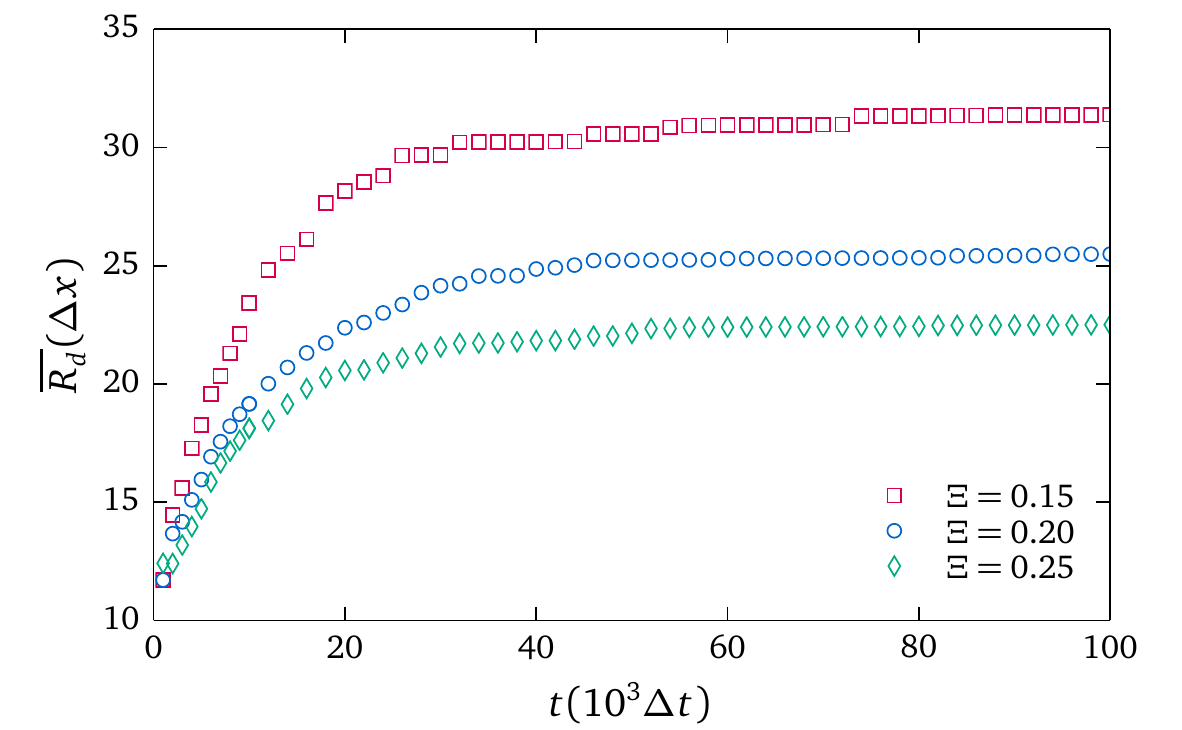}
  \caption{(Colour online) Time evolution of the surface weighted average droplet radius $\overline{\dr}$ of a Pickering emulsion, following Arditty \etal.~\cite{ bib:arditty-whitby-binks-schmitt-lealcalderon:2003}. We use the cluster sizes obtained through the Hoshen-Kopelman algorithm to calculate the radii $\dr$ of the droplets. The same qualitative behaviour as in Arditty \etal. Fig.~4 and Fig.~5 is observed: the droplet growth is rapid at first, but slows down until $\overline{\dr}$ converges to an asymptotic value.}
  \label{fig:arditty-time}
\end{figure}

\begin{figure}
  \cfig{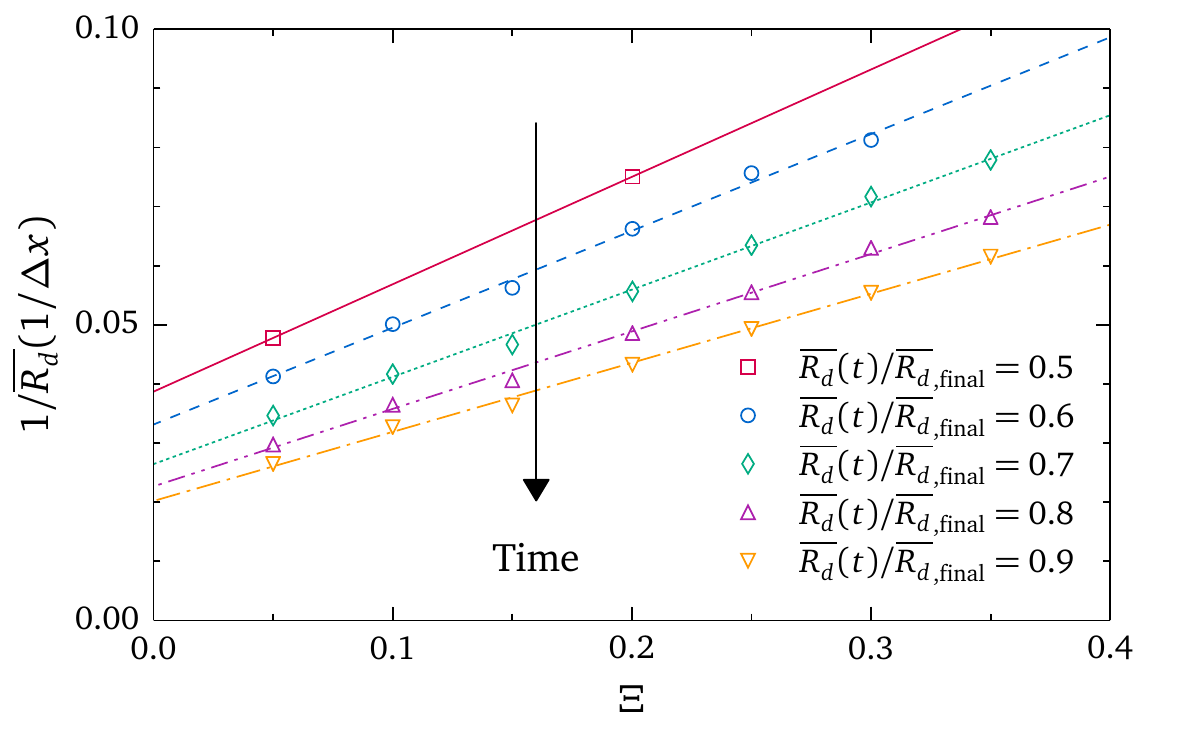}
  \caption{(Colour online) Dependence of the surface weighted average droplet radius $\overline{\dr}$ of a Pickering emulsion on the particle volume fraction $\pvf$, following Arditty \etal.~\cite{ bib:arditty-whitby-binks-schmitt-lealcalderon:2003}. We observe a linear relation with an offset; the offset can be explained by the facts that the system is finite, and that the interfaces are diffuse.}
  \label{fig:arditty-linear}
\end{figure}

We now consider emulsions for a range of particle volume fractions $0.05 \le \pvf \le 0.35$ and particle contact angles $50^\circ < \pca < 130^\circ$, in a system of volume $\svol = 256^3$ that has been allowed to equilibrate. The time allowed for equilibration is $t = 10^5$ time steps; we based the length of this interval on the time evolution of the domain sizes (discussed below), combined with visual inspection of the system. The first quantity to be considered is the average domain size, since its dependence on $\pvf$ is of particular interest. Qualitatively, increasing the number of particles in an emulsion will increase the interfacial area they can collectively stabilize, which then increases the surface-to-volume ratio of the fluid phases and makes the domains smaller on average. For a more quantitative analysis we follow Arditty \etal.~\cite{ bib:arditty-whitby-binks-schmitt-lealcalderon:2003} and define the surface weighted average droplet radius as
\begin{equation}
  \label{eq:swadr}
  \overline{\dr}(\time) = \frac{ \sum_i n_i (\dr)_i^3}{ \sum_i n_i (\dr)_i^2}
  \hbox{,}
\end{equation}
where $n_i$ is the number of droplets with radius $\dr$. In Fig.~\ref{fig:arditty-time} we show that our simulated emulsions display the same time dependence as the experimental emulsions discussed in Arditty \etal.: the droplet growth is rapid at first, but slows down until $\overline{\dr}$ converges to an asymptotic value. We have used the Hoshen-Kopelman method to calculate the droplet sizes (the HK algorithm supplies an average volume, which we represent by an average radius of a spherical domain of equal volume). Following the geometric arguments presented in~\cite{bib:arditty-whitby-binks-schmitt-lealcalderon:2003}, the surface weighted average droplet radius is expected to be inversely proportional to the total amount of solid particles present in the system (we use the particle volume fraction to quantify this, instead of the total mass of particles, which is used by Arditty \etal.):
\begin{equation}
  \label{eq:swadr-pvf}
  \frac{1}{\overline{\dr}(\time)} \sim \pvf
  \hbox{.}
\end{equation}
However, as is shown in Fig.~\ref{fig:arditty-linear}, we find an offset for
this linear relation, instead of it passing through the origin. The datasets
are based on the progression of the droplet sizes towards their long-time value
$\overline{\dr}_\mathrm{, final}$, where the rate of further growth cannot be
resolved anymore by the simulation. As we have seen that the droplet size
increases monotonically with time, the order of the datasets represents time
evolution. The existence of an offset can be attributed to two factors:
Firstly, the system is finite, and, even when the volume concentration of
particles drops, the droplets cannot grow infinitely large. Secondly, the
diffuse interfaces of the Shan-Chen model affect the droplet radii we measure,
but this is a constant contribution based on the thickness of the interfacial
region and thus is not taken into account in the scaling law.

\begin{figure}
  \cfig{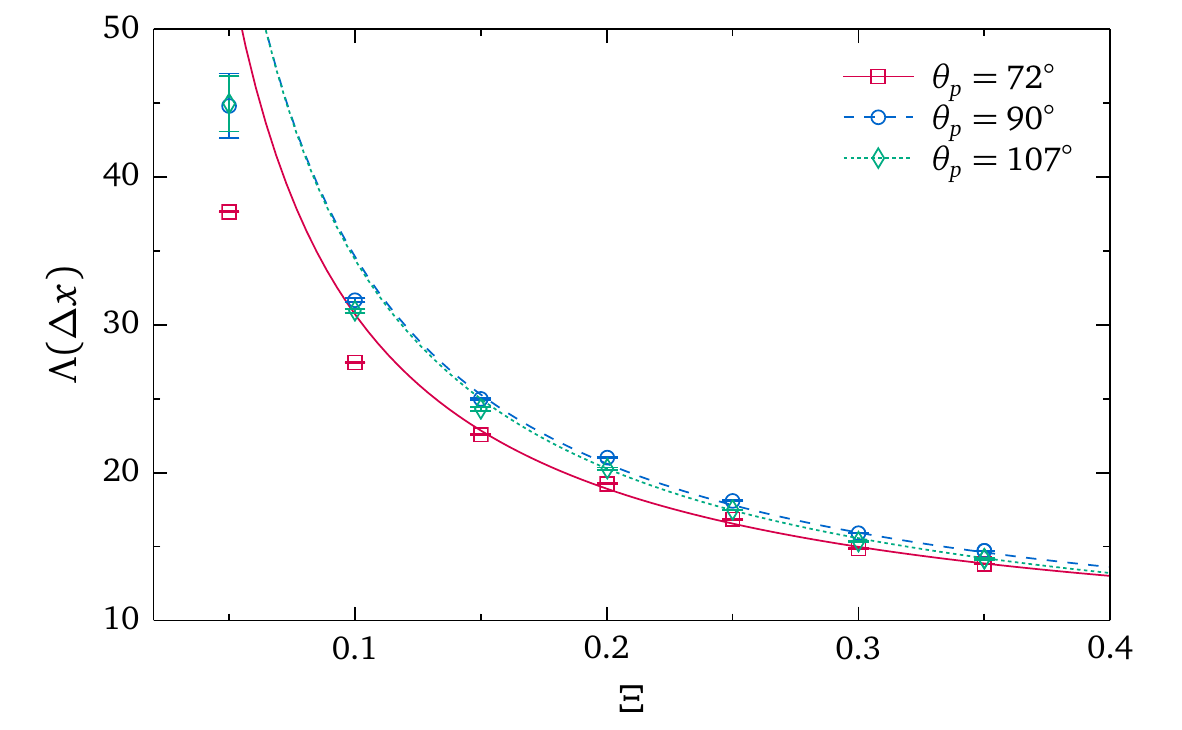}
  \caption{(Colour online) Averaged lateral domain size $\alds$ as a function of particle volume fraction $\pvf$ for various particle contact angles $\pca$. These values are then fit using an inverse relation with an offset, following Arditty \etal.~\cite{ bib:arditty-whitby-binks-schmitt-lealcalderon:2003} as in Jansen and Harting~\cite{ bib:jansen-harting:2011} resulting in $\alds(\pvf) = 7.11 + 2.36 / \pvf$ for $\pca= 72^\circ$ (solid line), $\alds(\pvf) = 6.61 + 2.80 / \pvf$ for $\pca = 90^\circ$ (dashed line), and $\alds(\pvf) = 6.14 + 2.83 / \pvf$ for $\pca = 107^\circ$ (dotted line). The domain sizes are largest in the case of neutrally wetting particles.}
  \label{fig:porous-emulsion-sf-pvf}
\end{figure}

\begin{figure}
  \cfig{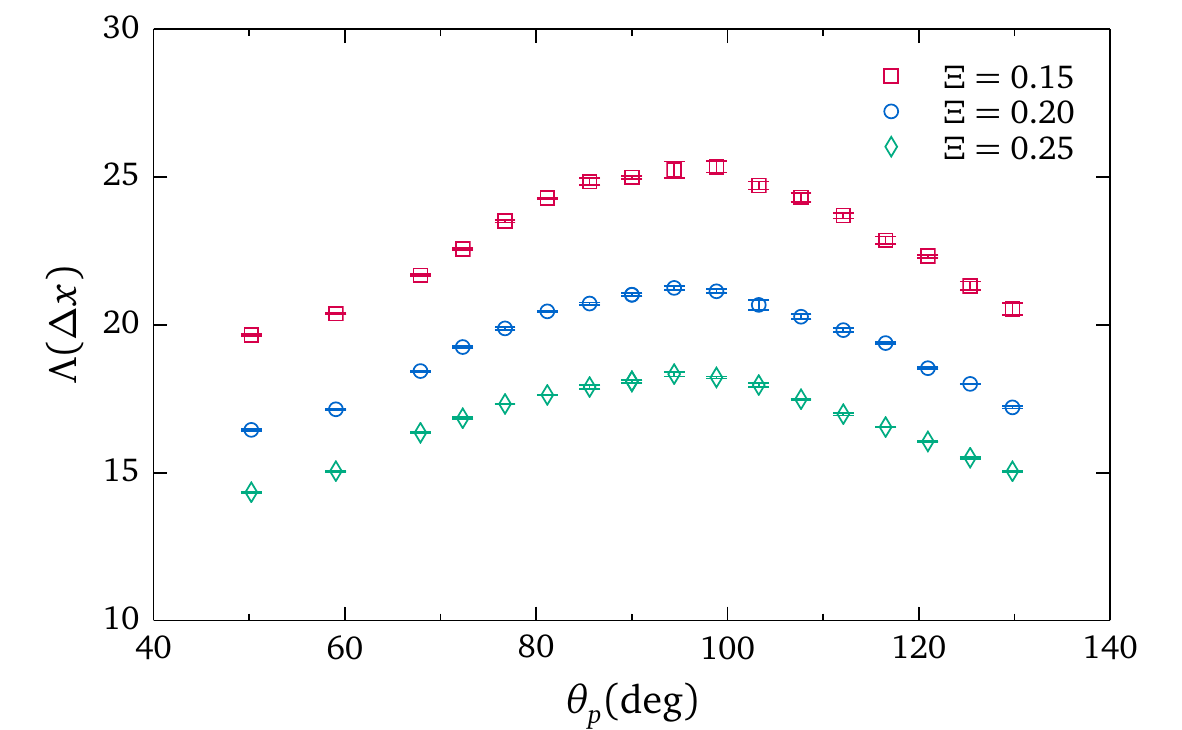}
  \caption{(Colour online) Average lateral domain size $\alds$ as a function of particle contact angle $\pca$ for various particle volume fractions $\pvf$. The domain size is largest for neutrally wetting particles. The interfacial curvature induced by the particles leads to smaller droplet sizes for Pickering emulsions. The reduction of $\alds$ as the contact angle deviates far from $90^\circ$ can be explained by the fact that particles have detached from the interface and moved into the bulk. They are then detected as domains of size $2 \pr = 10.0$, which depresses the average domain size towards that value.}
  \label{fig:porous-emulsion-sf-pca}
\end{figure}

\begin{figure}
  \centering
  \begin{tabular}{l l}
  \includegraphics[width=0.45\linewidth]{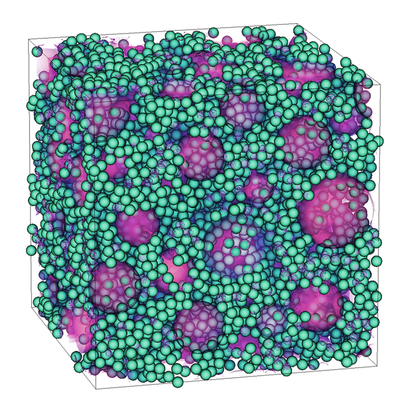} &
  \includegraphics[width=0.45\linewidth]{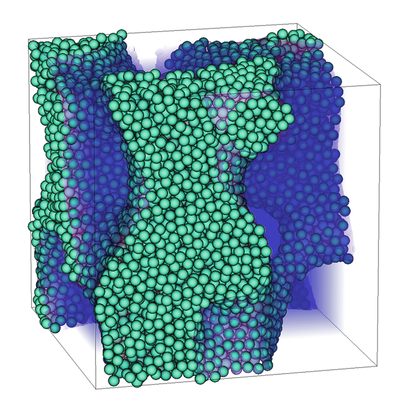} \\
  \end{tabular}
  \caption{(Colour online) Examples of degenerate emulsions: When the particles are strongly hydrophilic (here: $\pca = 50^\circ$) or hydrophobic (here: $\pca = 130^\circ$), they lose their ability to stabilize the system as their detachment energy decreases. In extreme cases, most particles will have moved into their preferred fluid, leaving a very small effective number of stabilizing particles at the interfaces. This strongly increases the average domain sizes, and the resulting state can hardly be called an emulsion at all.}
  \label{fig:emulsion-states-degenerate}
\end{figure}

These arguments also explain the data presented in Fig.~\ref{fig:porous-emulsion-sf-pvf}, which follows the way the data on domain sizes in emulsions is presented in Jansen and Harting~\cite{ bib:jansen-harting:2011}. Note that we follow their lead and now again use structure functions to calcluate domain sizes. We expect a relation of the form $\alds(\pvf) = a + b / \pvf$, with $a$ and $b$ fitting parameters. The offset $a$ is again a deviation from a purely inverse dependence law, and is caused by the same arguments as discussed above. This fitting function has been used to good effect in Fig.~\ref{fig:porous-emulsion-sf-pvf}, where $\alds(\pvf) = 7.11 + 2.36 / \pvf$ has been found for $\pca = 72^\circ$. For low particle volume fractions $\pvf < 0.15$, the interfacial area that can be stabilized gets so small that the domain sizes are even more strongly affected by the finite size of the system; as such, these data points are not used for fitting. The fit takes into account the error bars of the data points, which are the standard deviations of the components $\alds_i$ of the domain size. Surprisingly, this prodecure also produces satisfactory results for more bijel-like states based on $\pca = 90^\circ$ and $\pca = 107^\circ$: these can be fit well by $\alds(\pvf) = 6.61 + 2.80 / \pvf$ and $\alds(\pvf) = 6.14 + 2.83 / \pvf$, respectively. This method likely works because of the way the domain sizes are calculated (along Cartesian directions only), which in effect leads to quasi-ellipsoidal domains being measured; these then correspond to similar droplets as they would in the Pickering regime. Values for a wider range of contact angle have been calculated but are not shown here, as they follow the same trends. As expected, the change of contact angle only weakly changes the interfacial area that can be stabilized, and as such, there is only a weak dependence of the domain sizes on it.

Note however that largest values of $\alds$ are not encountered at either extreme of the range of contact angles, but instead in the middle: this is shown in more detail in Fig.~\ref{fig:porous-emulsion-sf-pca}, which clearly demonstrates a non-monotonic dependence of $\alds$ on $\pca$, and extends the range of contact angles in both directions as compared to Fig.~\ref{fig:porous-emulsion-sf-pvf}. Recall that the irreversibility of the adsorption of particles to a fluid-fluid interface is based on the large energy cost associated with the fluid-fluid interface that would exist in absence of the particle, and that this quantity depends linearly on both the surface tension and the area of the covered interface. As particles gain a preference to reside in one fluid, they will initially still be attached to the interface, but the area they cover will decrease: this follows from the simple geometrical argument that a planar cut through a sphere has the largest area if it goes through the centre of the sphere. As the particle centre moves away from the interface, the associated decrease in interface covering will decrease the detachment energy keeping the particle adsorped~\cite{ bib:binks-horozov:2006}:
\begin{equation}
  \label{eq:detachment}
  \Delta E_{\mathrm{det}} = \pi \pr^2 \st \left( 1 - |\cos \pca | \right)^2
  \mbox{.}
\end{equation}
The fluid-particle interface will also become more favourable, as long as the particle moves into its preferred fluid. Both effects combine to destroy the irreversibility of the particle adsorption, which strongly decreases the number of particles actually participating in interface stabilization. The energy barrier does not have to be of the order of thermal fluctuations: when the emulsion is formed the absence of interfaces and inter-particle collisions will keep particles from adsorbing in the first place. One would expect this effect to cause an increase of the average domain size, however, Fig.~\ref{fig:porous-emulsion-sf-pca} shows the opposite. This is caused by the way the structure factor is calculated: it measures fluctuations in the order parameter and the simulation code returns a zero value for all sites occupied by a particle. If this particle is located at the interface, where the order parameter of the fluids is close to zero, it will not add a strong signal because it blends into the fluid background. However, any particles that are inside one of the fluid phases, where the order parameter is far away from zero, will add a signal to the calculation, which adds a length scale comparable to their diameter. Because the droplet diameters are larger than the particle diameters, detached particles artificially reduce the value of $\alds$. Aside from distorting the measured domain sizes, the emulsions stabilized by particles with extreme contact angles present very differently from the variants with more moderate contact angles; examples are shown in Fig.~\ref{fig:emulsion-states-degenerate}. The domains are much coarser, and these degenerate emulsions can hardly be called emulsions at all. Note that in the case of a Pickering emulsion, the individual droplets are no longer covered by a densly packed monolayer; instead the droplets are separated by clumps of particles. This strongly reduces the long-term stability of the state as Ostwald ripening is no longer effectively blocked.

\begin{figure}
  \cfig{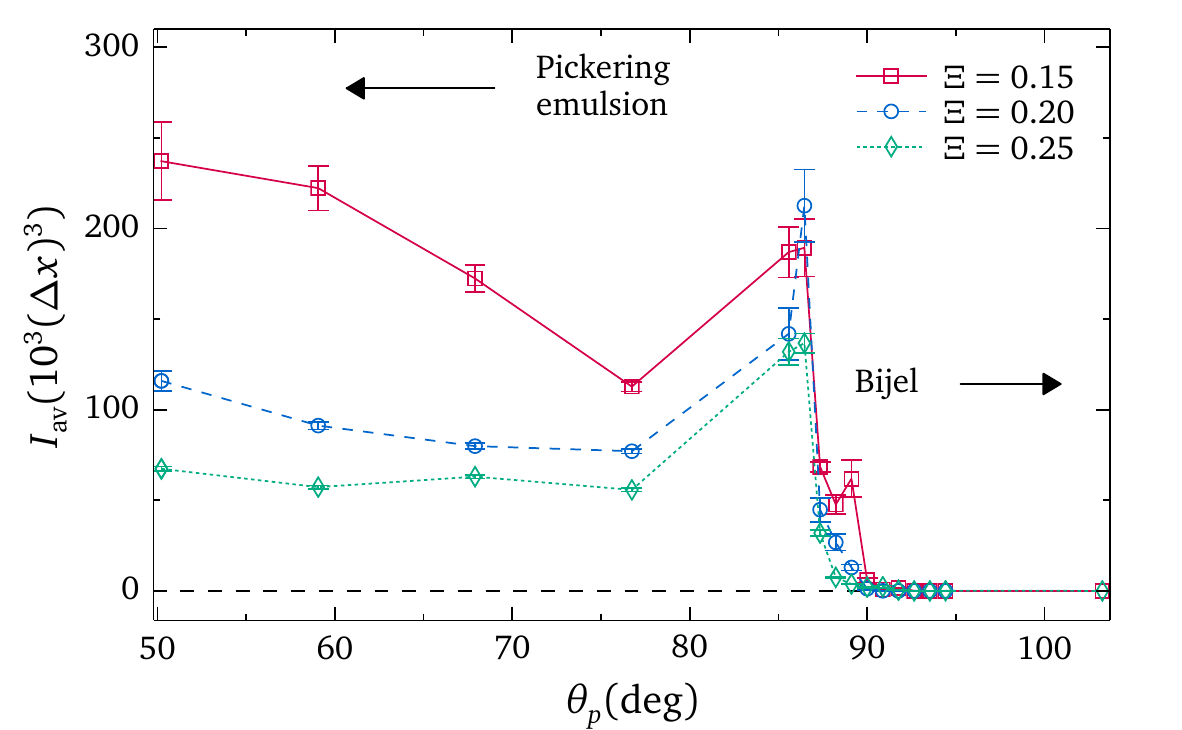}
  \caption{(Colour online) Reduced average cluster size $I_{\mathrm{av}}$ after equilibration of $\time = 10^5$ time steps as a function of the contact angle $\pca$ of the stabilizing particles. After subtracting the effect of the largest cluster, the bijel has zero average cluster size, while the Pickering emulsions have nonzero sizes, with a dependence on the particle volume fraction $\pvf$. For all data points, the system size is $\svol = 256^3$ lattice sites, the fluid-fluid ratio $\ffr = 0.56$ ($\den^r_{\mathrm{init}} = 0.5$ and $\den^b_{\mathrm{init}} = 0.9$), and the surface tension is $\st = 0.014$. The error bars are calculated through the standard deviation of the droplet sizes.}
  \label{fig:racs-initial}
\end{figure}

\Cref{fig:racs-initial} also depicts the dependence of domain sizes on the contact angle of the particles, but here we use the reduced average cluster size (this corresponds to a volume, not a length). The error bars are calculated through the standard deviation of the droplet sizes (again excluding the largest droplet). We clearly see the advantage of this method when attempting to detect a transition from a bijel to a Pickering emulsion or vice versa: $I_{\mathrm{av}}$ sharply drops to zero when a bijel state is achieved. In the Pickering regime, the average domain size is a measure for the droplet sizes, while for a bijel $I_{\mathrm{av}} = 0$. In the transition region a peak can occur: this corresponds to multiple bijel-like domains co-existing, only one of which is discarded during the calculation. This leaves one with one or more very large clusters which strongly contribute to raising the average (the number of discrete droplets will be small at this point). As this method is not affected by the presence of detached particles (except to subtract their volume from the domain size), we now properly observe the increase of the average domain size as the particle contact angle is reduced and the particles are more hydrophilic.

\begin{figure}
  \cfig{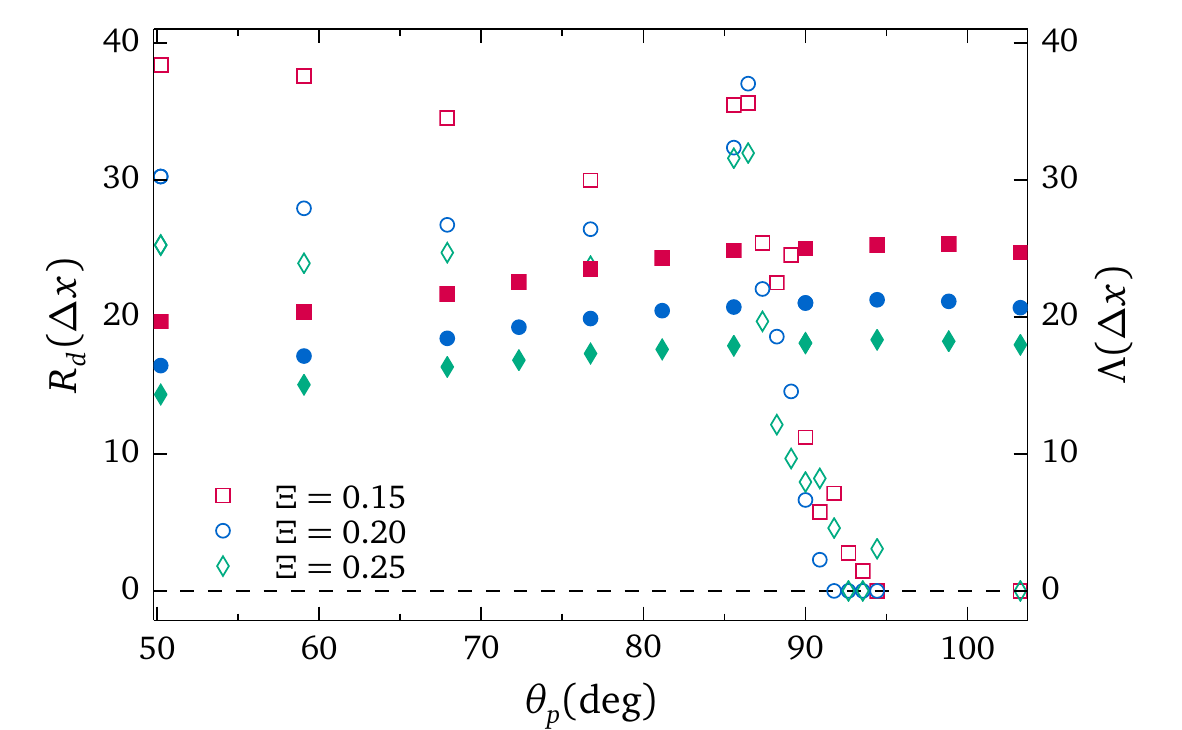}
  \caption{(Colour online) We compare the values of the reduced average cluster size (shown as radius $\dr$, open symbols) to those of the averaged lateral domain size $\alds$ (filled symbols) for the same datasets. Very large differences exist: in the Pickering regime the value of $\alds$ is depressed because the method detects the particle size and inter-particle distances as characteristic length scales, while the cluster sizes are calculated based on the number of lattice sites of the fluid domains. In the bijel regime, the cluster size of the single domain in a bijel will not provide meaningful information on the size of local structures.}
  \label{fig:racs-R-initial}
\end{figure}

Throughout this section we have alternately used two methods to measure the number of lattice sites of domains. To conclude this section, we show both the averaged lateral domain size $\alds$ (filled symbols) and the reduced average cluster size (transformed into a radius $\dr$, shown as open symbols) as a function of the particle contact angle $\pca$ (cf. Fig.~\ref{fig:racs-R-initial}), effectively combining Fig.~\ref{fig:porous-emulsion-sf-pca} and Fig.~\ref{fig:racs-initial}. The reduced average cluster size is again represented by an average radius of spherical domains of equal volume, for a more convenient comparison. We see that although both methods have their own advantages and disadvantages such that different situations may warrant the use of one method over the other, care should be taken when attempting to compare results directly: in the Pickering regime the value of $\alds$ is depressed because the method detects the particle size and inter-particle distances as characteristic length scales, while the cluster sizes are calculated based on the number of sites of the fluid domains. In the bijel regime, the cluster size of the single domain in a bijel will not provide meaningful information on the size of local structures.

\subsection{Droplet size distribution}

One clear advantage of the Hoshen-Kopelman method is the possibility to obtain data on individual cluster sizes, and not only on averages. In this section we briefly discuss the distribution of droplet sizes in Pickering emulsions. Similarly to the data presented above, we introduce a cutoff to eliminate non-physical domains, and present the data terms of the radii $\dr$ of spherical domains that have the same volume as the measured clusters. 

\begin{figure}
  \cfig{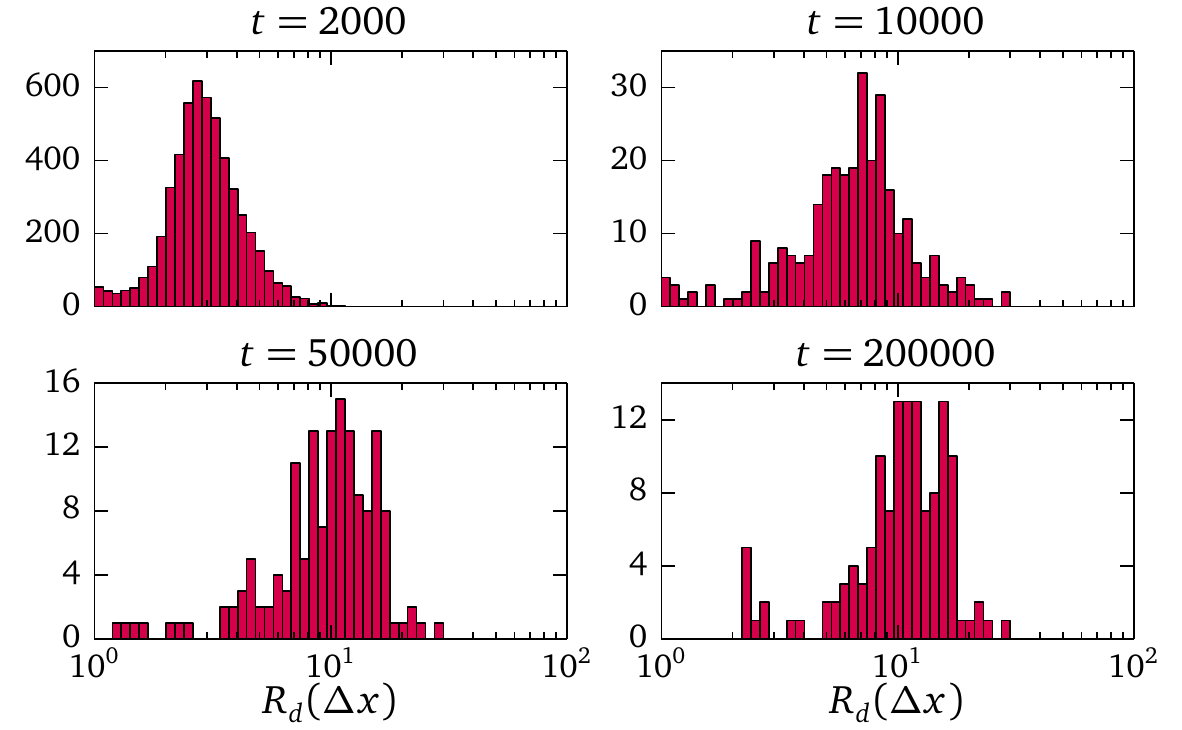}
  \caption{(Colour online) Following Arditty \etal.~\cite{ bib:arditty-whitby-binks-schmitt-lealcalderon:2003} we present our data on a logarithmic scale for the droplet sizes. It is evident that after the early time evolution, when the droplets are small but numerous and the resulting distribution is smooth, the statistics of even five combined Pickering emulsions of a $256^3$ system leave to be desired (note the change in range of the vertical axes). Acquiring the necessary statistics to smooth out the latter distributions is computationally prohibitive.}
  \label{fig:droplet-stacked-histogram-log}
\end{figure}

The size of the stabilised droplets is often of interest, and not only as an
average quantity. For example, it will strongly affect the mouthfeel of
particle-stabilised food products. As such, we are interested in tracking the
time evolution of the droplet size distribution in a Pickering emulsion,
starting from a random mixture of fluids and particles. The results of such a
procedure can be seen in Fig.~\ref{fig:droplet-stacked-histogram-log}. The bars
of the histograms comprise the combined data of five systems with identical
physical properties but different random initial distribution of the fluids and
particles.

Three regimes have been identified in the time evolution of a Pickering
emulsion~\cite{ bib:guenther-frijters-harting:2014}, and we discuss an example
here. Firstly, after we have initialised the simulation, demixing quickly sets
in. This is caused by the immiscibility of the fluids, and effects formation
of droplets (corresponding to the formation of many small domains in terms of
the HK algorithm). After $t = 2000$ LB time steps the distribution of
droplet radii is narrow and centered around a radius $\dr \approx 3$. The radii
are of the order of the interfacial thickness of the Shan-Chen multicomponent
model and as such we do not expect them to be quantitatively accurate; however,
their behaviour is correct in a qualitative sense. The droplets then continue
to grow rapidly due to the process of Ostwald ripening; this leads to both
larger average droplet radii and a wider distribution because the system
becomes less homogeneous. Furthermore, the total number of droplets decreases
rapidly. This process continues for some tens of thousands of time steps; then,
Ostwald ripening is halted due to the almost complete covering of fluid-fluid
interfaces by particles. The droplets can grow by coalescence only, which is a
rare phenomenon due to the particle layers and long diffusion times. The
average droplet radius remains constant at $\dr \approx 10$. We show that the
distribution of the droplet sizes does not change significantly after $5 \cdot
10^4$ time steps. The observed monodispersity of the droplets, the shift of the
position of the peak in time, as well as the (slight) increase in uniformity of
the droplet sizes (which manifests iself as a decrease of the width of the
distribution relative to the position of the peak) qualitatively matches the
behaviour shown in Arditty \etal.~\cite{
bib:arditty-whitby-binks-schmitt-lealcalderon:2003}. However, quantitative
agreement has proven difficult to obtain: even when the results of multiple
simulations are combined, the accuracy of the statistics obtained from those
simulations is limited, as we show in
Fig.~\ref{fig:droplet-stacked-histogram-log}. After the early time evolution, when the
droplets are small but numerous and the resulting distribution is smooth, the
statistics of even five combined Pickering emulsions of a $256^3$ system are
insufficient to provide reliable data (note the change in range of the vertical
axes). Acquiring the necessary volume of data to smooth out the latter
distributions is computationally prohibitive. We have not studied this subject
further in any great detail, but the availability of even this type of limited
data on the distribution of the droplet sizes can help to, e.g. identify
parameters to help create monodisperse droplets of optimal size.

\section{Conclusion}
\label{sec:emulsions-conclusions}
In this paper we presented computer simulations of particle-stabilized emulsions consisting of two immiscible fluids and colloidal particles. The final state an originally random mixture of fluids and particles reaches after phase seperation can be classified as a Pickering emulsion, a bijel, or an intermediate state. Which state the system reaches depends on a number of parameters, such as the ratio between the two fluids, the particle volume concentration, and the particle contact angle. One of the emulsion properties of interest is the size of the droplets in the case of Pickering emulsions, and the length scales of local structures, in the case of bijels. We have characterized these using the three-dimensional structure function, as well as the sizes of connected domains, using a novel parallelized implementation of the Hoshen-Kopelman algorithm~\cite{ bib:frijters-krueger-harting:2014}. The former method provides useful information in the case of Pickering emulsions as well as bijels, where the latter does not provide useful data for bijels. However, it does provide information on individual droplet sizes, rather than just averages. This feature can be used to study the size distribution of the droplets.

We have followed the work of Arditty \etal.~\cite{
bib:arditty-whitby-binks-schmitt-lealcalderon:2003} and confirmed an inverse
dependence between the concentration of particles and the average radius of the
stabilized droplets. In our simulations, the finite system and diffuse
interface of the Shan-Chen model cause an offset to this law. We have also
observed that after a rapid growth of droplets at the start (nucleation), the
growth slows down (Ostwald ripening) and eventually halts (coalescence of
droplets only). This analysis follows Jansen and Harting~\cite{
bib:jansen-harting:2011}, but uses the HK algorithm to obtain droplet sizes and
adds a substantially deeper insight and more thorough variation of parameters.
Care has to be taken when comparing results from the structure factor and the
HK algorithm: although they attempt to describe the same physical quantities
they both have their weaknesses. The structure factor interprets the presence
of detached particles as domains, which artificially reduces the measured
average domain size. This is especially problematic when particle contact
angles become extreme, and degenerate bijels or Pickering emulsions are formed.
The HK algorithm just counts the number of sites in a fluid domain and is thus
not able to provide any meaningful information about the structure of domains
in the bijel regime. 

Finally, the Hoshen-Kopelman has been used for a brief look at the distribution
of individual droplet sizes in Pickering emulsions: the results qualitatively
match previous experimental observations.


\begin{acknowledgments}
Financial support is acknowledged from the FOM/Shell IPP (09iPOG14 -
``Detection and guidance of nanoparticles for enhanced oil recovery'') and
NWO/STW (Vidi grant 10787 of J.~Harting). We thank the J\"ulich Supercomputing
Centre and the High Performance Computing Center Stuttgart for the technical
support and the CPU time which was allocated within large scale grants of the
Gauss Center for Supercomputing.
\end{acknowledgments}

\end{document}